\documentclass[prl, superscriptaddress, twocolumn]{revtex4-1}

\usepackage{amssymb,amsmath}
\usepackage{latexsym}
\usepackage{graphicx}
\usepackage{epstopdf}

\usepackage{color}
\usepackage{graphics}
\usepackage{epsfig,subfigure}
\usepackage{bm}

\newcounter{mnotecount}[section]
\renewcommand{\themnotecount}{\thesection.\arabic{mnotecount}}
\newcommand{\mnote}[1]
{\protect{\stepcounter{mnotecount}}$^{\mbox{\footnotesize  $
      \bullet$\themnotecount}}$ \marginpar{\raggedright\tiny
    $\!\!\!\!\!\!\,\bullet$\themnotecount: #1} }

\begin{document}

\newcommand{\dR}{\mathbb R}
\newcommand{\dC}{\mathbb C}
\newcommand{\dZ}{\mathbb Z}
\newcommand{\id}{\mathbb I}
\newtheorem{theorem}{Theorem}
\newcommand{\ud}{\mathrm{d}}
\newcommand{\mfn}{\mathfrak{n}}

\author{Herv\'{e} Bergeron}
\affiliation{ISMO, UMR 8214 CNRS, Univ Paris-Sud, France}

\author{Ewa Czuchry}
\affiliation{National Centre for Nuclear Research, 00-681
Warszawa, Poland}

\author{Jean-Pierre Gazeau}
\affiliation{APC, Universit\'e Paris Diderot, Sorbonne Paris Cit\'e, 75205 Paris
Cedex 13, France}
\affiliation{Centro Brasileiro de Pesquisas Fisicas
22290-180 - Rio de Janeiro, RJ, Brazil }

\author{Przemys{\l}aw Ma{\l}kiewicz}
\affiliation{National Centre for Nuclear Research,  00-681
Warszawa, Poland}
\affiliation{APC, Universit\'e Paris Diderot, Sorbonne Paris Cit\'e, 75205 Paris
Cedex 13, France}

\author{W{\l}odzimierz Piechocki}
\affiliation{National Centre for Nuclear Research,  00-681
Warszawa, Poland}

\date{\today}

\title{\bf Smooth Quantum Dynamics of Mixmaster Universe}

\begin{abstract}
We present a quantum version of the vacuum Bianchi IX model by
implementing affine coherent state quantization  combined with a
Born-Oppenheimer-like adiabatic approximation. The  analytical
treatment is carried out on both  quantum and semiclassical
levels. The resolution of the classical singularity occurs by
means of a repulsive potential generated by our quantization
procedure.   The quantization of the oscillatory degrees of
freedom  produces a radiation energy density term in the
semiclassical constraint equation. The Friedmann-like lowest
energy eigenstates of the system are found to be dynamically stable.

\end{abstract}
 \maketitle

\noindent{\it  Introduction.}  The Friedmann-Robertson-Walker
model is successfully used to describe the data of observational
cosmology (see e.g. \cite{Green:2014aga,Ade:2013zuv}). 
Nevertheless, the isotropy of space is dynamically unstable towards the big-bang singularity \cite{Lif}.  
On the other hand, if the present Universe  originated from an inflationary phase, then the pre-inflationary universe is supposed to have been both inhomogeneous and anisotropic. As evidence suggests   (see \cite{BKL1}, \cite{Gar}), the  dynamics of such universe backwards in time becomes ultralocal and  
effectively identical with the homogeneous but anisotropic one at each spatial point.
In both cases  quantization of the isotropic models  alone appears to be insufficient.
 Hence the quantum version of an  anisotropic model,    comprising the Friedmann model as a particular case,  is expected to be better suited for describing the earliest Universe.

In the Letter we advocate a new quantization method of the
 dynamics of a vacuum Bianchi type IX geometry, the Mixmaster universe.
 We identify a soluble sector of this model,
which lies deeply in the quantum domain and, as we show, contains  relevant
physics.

The Mixmaster universe exhibits a complex behavior \cite{cwm}.  As it  
collapses, the universe enters
chaotic oscillations producing an infinite sequence of
distortions from its spherical shape \cite{NJC}. Those distortions essentially correspond to the level of anisotropy and may be viewed as
an effect of a gravitational wave evolving in an isotropic
background  \cite{DHK}. Dynamics of this wave is nonlinear, and its
interaction with the isotropic background fuels the gravitational
contraction. Not surprisingly, the quantization of Bianchi  IX model is a
difficult task. {Some formulations}  can be
found in literature, { including} the Wheeler-DeWitt equation
\cite{cwm} or, more recently, a formulation based on loop
quantum cosmology \cite{Bojowald:2003xe,WilsonEwing:2010rh}.
However, the search for solutions within these formulations is
quite challenging \cite{Marolf, Moncrief} leaving the near big-bang dynamics largely unexplored. The most recent developments,  e.g. \cite{Bae}, do not address the singularity resolution.

To fill this gap we propose a quantum Mixmaster dynamics, which
originates from affine coherent state (ACS) quantization  that
was recently used to obtain the quantum  Friedmann model
\cite{Bergeron:2013ika}. { It was shown   that ACS quantization causes some new terms to appear in the quantum Hamiltonian,  producing a strong
repulsive force counteracting the contraction of universe.} The
capacity to resolve the singularity constitutes the basic
advantage of our quantization method. In order to solve the
dynamics in the present, more complex setting, we employ the
adiabatic approximation widely utilized in quantum molecular
physics \cite{born51,sutcliffe12}. This approach is reasonable
when  the vibrations of the shape of the universe are
significantly faster than the contraction  of its volume.

The main result is a semiclassical Friedmann-like equation
obtained from expectation values in ACS, {a description  peculiar to our approach.} In that equation, the
expansion of the universe is governed by two terms of quantum
origin. The first one is proper to the quantum Mixmaster model and
corresponds to the energy of the wave in an eigenstate. It is
proportional to the energy level number. The other one, which is
more universal, corresponds to the repulsive potential preventing
the singularity. The lowest energy eigenstates of this system are
interpreted as the quantum Friedmann universe supplemented with
vacuum fluctuations of the anisotropy.

Beyond issues of singularity resolution, the Friedmann-like equation
describes two novel and rather surprising properties of the
quantum dynamics. Firstly, the anisotropic degrees of freedom
remain in their lowest energy states during the quantum phase
consistent with our approximation. This implies that the quantum
Friedmann model, unlike its classical counterpart, is in fact
stable with respect to the anisotropy. Therefore, the classical
chaos is suppressed within the considered domain. Secondly, during
the contraction the quantum energy of anisotropy grows much slower
than it does on the classical level. Namely, it effectively
gravitates as radiation { leading to} a significant
reduction in the overall gravitational pull from anisotropy due to
quantum effects.

\noindent{\it  Classical Hamiltonian.} The Friedmann equation
extended to anisotropic universes, with $c=1=8\pi G$, reads:
\begin{equation}\label{friedmann}
H^2+\frac{1}{6} ~^3R-\frac{1}{6}\Sigma^2=\frac{1}{3}\rho\,,
\end{equation}
where $H=\dot{a}/a$  (dot denotes the derivative with respect to
the cosmological time) is the expansion parameter,  $\rho$ is the
energy density of matter, and $^3R$ is the spatial curvature. The
additional term $\Sigma^2$ is the total shear of the spatial
section and is non-vanishing for anisotropic models. Due to its
negative sign, the shear drives the gravitational collapse and it
eventually dominates the dynamics. For this reason we neglect the
attraction of matter by putting $\rho=0$, that is, we restrict our considerations to the vacuum Bianchi IX, whose dynamics near the singularity is effectively the same as in the presence of perfect fluid \cite{BKL1}.

The Mixmaster describes the space-time metric $
ds^2=-dt^2+a^2(e^{2\beta})_{ij}\sigma^i\sigma^j
$, where $a$ is an averaged scale factor and  $\sigma^i$ are  differential forms on a three-sphere (covering group of the rotation group) satisfying $d\sigma^i=\frac12 \epsilon_{ijk} \sigma^j\wedge\sigma^j$. The diagonal form of the metric is assumed $(e^{2\beta})_{ij}:= \mathrm{diag} \;(e^{2 ( \beta_+ + \sqrt{3} \beta_-)},
e^{2 (  \beta_+ - \sqrt{3} \beta_-)}, e^{ 4\beta_+})$, where   $\beta_{\pm}$ are distortions parameters \cite{cwm}.

\begin{figure}[!t]
\includegraphics[width=6cm]{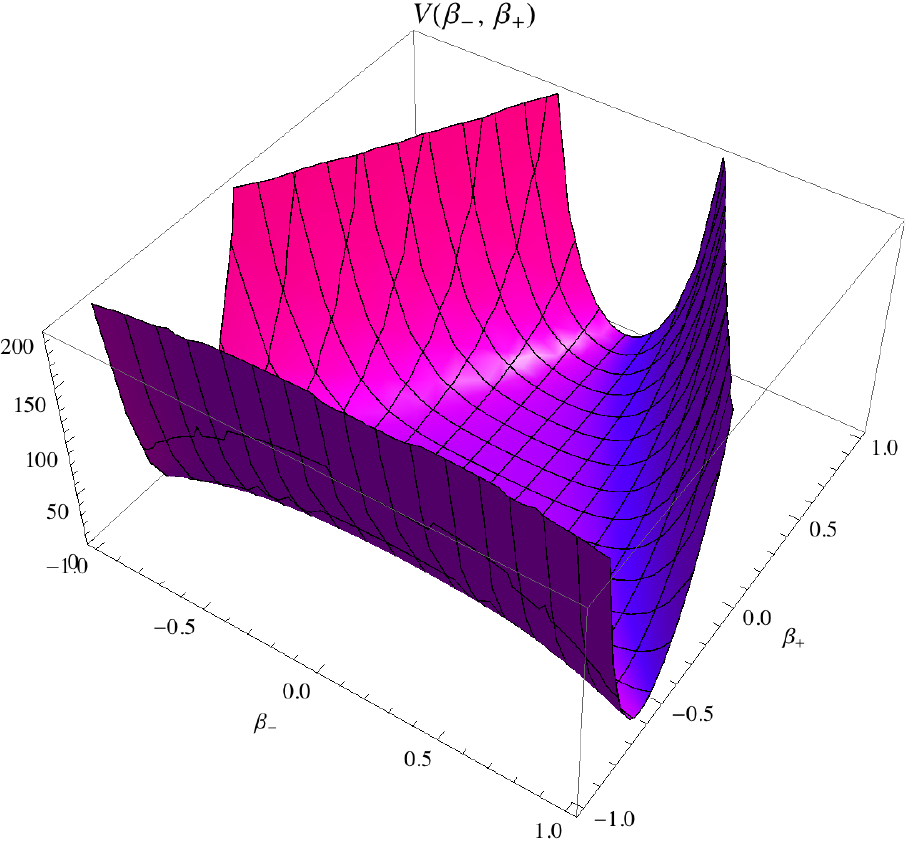}
\caption{Global picture of the potential $V(\beta) $ near its
minimum. Boundedness from below,  confining aspects, and three
canyons are illustrated.} \label{figure2}
\end{figure}

In terms of these variables, the shear is the kinetic energy of anisotropic distortion:
$\Sigma^2=(p_+^2+p_-^2)/24a^6$, where the momenta
$p_{\pm}$ are canonical conjugates to 
$\beta_{\pm}$. The spatial curvature $^3R$ grows due to the overall contraction of space, but decreases due to the growth of
anisotropy. This last circumstance leads to the backreaction from
spatial curvature on the shear and, as a result, oscillations in
$\beta_{\pm}$ occur. As there is no matter content in our model,
$\beta_{\pm}$ describe a sort of gravitational wave. The curvature
can be split into isotropic  and anisotropic parts:
$~^3R=3(1-V(\beta))/2a^2$, where $V(\beta)$ is
the anisotropy curvature potential \cite{cwm}:
\begin{equation*}\label{h2}
V(\beta) =  \frac{e^{4 \beta_+}}{3} \left( \left( e^{-6\beta_+} -
2 \cosh (2 \sqrt{3} \beta_-) \right)^2 - 4 \right)+1.
\end{equation*}
As shown in Fig.~\ref{figure2}, this potential has three ``open''
${\sf C}_{3v}$ symmetry
directions. One can view them as three deep ``canyons'',
increasingly narrow until their respective wall edges close up at
the infinity whereas their respective bottoms tend to zero.
Due to its  (almost) confining shape,
$V$ is expected to produce a discrete energy spectrum on the
quantum level.

The generalized Friedmann equation (\ref{friedmann})  may be
rewritten as
\begin{equation}\label{friedmann2}
H^2+\frac{1}{4 a^2}=\frac{1}{6}\Sigma^2-\frac{V(\beta)}{4 a^2}
\end{equation}
where the isotropic background geometry on the l.h.s. is pulled by
the energy of anisotropic oscillations. The energy of oscillations
scales roughly as $ a^{-6}$.

It follows that the Hamiltonian constraint to be quantized reads
in canonical variables:
\begin{align}\label{h4}
\mathcal{C}&=  \frac{3}{16 } p^2+\frac{3}{4 } q^{2/3}
-{\mathcal H}_q ,
\end{align}
where $q=a^{3/2}$ and $p^2=16\dot{a}^2a$ are more
suitable to ACS quantization, and where
\begin{align}
\label{h4a}{\mathcal H}_{q}  &= \frac{1}{12 q^2} (p_+^2+p_-^2)
+\frac{3}{4}q^{2/3} V(\beta)
\end{align}
is the anisotropy energy. The closed  Friedmann-Robertson-Walker (FRW) geometry is obtained by putting
$p_\pm=0$ and $\beta_\pm=0$, or simply ${\mathcal H}_q =0$.

The  Hamiltonian constraint \eqref{h4}  resembles a 
diatomic molecular Hamiltonian with the pairs $(q,p)$ and
$(\beta_\pm,p_\pm)$ playing the r\^ole of the reduced nuclear and
electronic  variables, respectively.
In molecules, the motion of nuclei
is slow enough in comparison with electrons so the  motion of electrons may
be approximated as becoming immediately adjusted to varying positions of nuclei.
However,  the coupling between the nucleus-like and electron-like degrees
of freedom in the present model differs from the usual molecular case for which the
validity of the approximation rests upon the smallness of the
ratio between the nuclei and electron masses. In the present case, described
by  Eqs.\;\eqref{h4}
and \eqref{h4a}, the ``mass'' of the degrees of freedom $\beta_\pm$ evolves as $
 q^2$. Thus, it goes to zero near the singularity, $q = 0$.
On the other hand, the ``mass'' of the degree of freedom $q$ in
Eq.\;\eqref{h4} is  constant. Thus, close to singularity the latter may be regarded
as ``heavy'' in comparison with the  anisotropic variables that can be treated as ``light''.

\noindent{\it  Quantum Hamiltonian.} The six-dimensional phase
space of the Mixmaster universe is quantized as follows: (A) The
isotropic variables form the canonical pair $(q,p)$ living in a
half-plane. That half-plane can be viewed as the affine group. We
resort to one of its two unitary irreducible representations, denoted by $U$, to build from a suitable fiducial vector
$|\nu\rangle$ (where $\nu >0$ is a free parameter) a family of
affine coherent states (i.e., wavelets) $|q,p\rangle :=
U(q,p)|\nu\rangle$. These ACS's are then used to consistently
quantize the isotropic variables. While the method provides the
usual $\hat{p}=-i \hbar
\partial_q$, and $\hat{q}$ defined as the multiplication by $q$,
its interest lies in the regularization of the Hamiltonian \cite{Bergeron2014}. This
approach together with $|\nu\rangle$ was introduced for cosmological models in
\cite{Bergeron:2013ika}. Next, we use the ACS's to obtain a
semiclassical description, which enables us to analyze the
effective dynamics of isotropic variables. (B)  For the
anisotropic variables, each canonical pair $(\beta_\pm, p_\pm)$
lives in the plane. Thus, it is natural to proceed with the usual
canonical quantization which yields $\hat{p}_\pm = - i \hbar
\partial_{\beta_\pm}$,
and $\hat{\beta}_\pm$ being the multiplication by $\beta_\pm$.

The quantized Hamiltonian corresponding to
\eqref{h4} and issued from quantizations (A) and (B) above reads
\begin{equation}
\label{hq} \hat{\mathcal{C}} =  \frac{3}{16} \left( \hat{p}^2 +
\frac{\hbar^2 \frak{K}_1}{\hat{q}^2} \right) +\frac{3}{4} \,
\frak{K}_3\, \hat{q}^{2/3} - \hat{\mathcal H}_q \,,
\end{equation}
where
\begin{equation}\label{h5}
\hat{\mathcal H}_q = \frac{1}{12} \frak{K}_2 \frac{
\hat{p}_+^2+\hat{p}_-^2}{\hat{q}^2} +\frac34 \frak{K}_3\, q^{2/3}
 V(\hat{\beta}) \,.
\end{equation}
The $\frak{K}_i$'s are purely positive
numerical constants dependent on the choice of the ACS. With the
choice made in our previous paper \cite{Bergeron:2013ika}  all
these constants  are simple rational  functions of modified Bessel
functions $K_l(\nu)$.
We note in (\ref{hq}) the appearance of the repulsive centrifugal
potential term $\hbar^2 \frak{K}_1 \hat{q}^{-2}$. It is the signature of the
ACS  quantization, which is consistent with the half-plane
geometry, and it regularizes the quantum Hamiltonian for small
$q$ \cite{Bergeron2014}. As the universe approaches the singularity,
$q\rightarrow 0$, this centrifugal term sharply grows in
dynamical significance.

We consider  the oscillations of $\beta_{\pm}$ fast in comparison
with the contraction of the universe. It legitimates the adiabatic
approximation, in a way analogous to the Born-Oppenheimer
approximation (BOA) \cite{born51,sutcliffe12} widely used in
molecular physics. Due to the confining property of $V$, the
operator $\hat{\mathcal H}_{q}$ at fixed $q$ has a
discrete spectrum. In accordance  with BOA we assume that the
anisotropy degrees of freedom $\beta_\pm$ are frozen in some
eigenstate of $\hat{\mathcal H}_q $ with eigenenergy
$e^{(N)}_{q}$ ($N=0,1,\dots$) evolving adiabatically.
Thus, the light degrees of freedom $\beta_\pm$  can be { averaged}
leading to the Hamiltonian:
\begin{equation}\label{h6}
\hat{\mathcal{C}}_{A} =  \frac{3}{16} \left( \hat{p}^2 +
\frac{\hbar^2 \frak{K}_1}{\hat{q}^2} \right)+\frac{3}{4}
\frak{K}_3 \hat{q}^{2/3} -  e^{(N)}_{q} \,.
\end{equation}
Focussing on the deep quantum domain, we look at the first energy levels near the ground state of $\hat{\mathcal H}_q $. Therefore we essentially investigate the domain near the minimum of $V(\hat{\beta})$. Within the harmonic approximation $V(\hat{\beta})\approx 8
(\hat{\beta}_+^2+\hat{\beta}_-^2) $ near its minimum
$\hat{\beta}_\pm=0$, the eigenenergies are found to be
$
e^{(N)}_{q} \simeq {\hbar }{{q}^{-2/3}} \sqrt{2 \frak{K}_2
\frak{K}_3} (N+1)
$ with $N= n_+ + n_-$. 
The quantum numbers $n_\pm$ correspond
to the independent harmonic oscillations in $\beta_+$ and
$\beta_-$. Consequently, the approximation of the eigenvalues
of $\hat{\mathcal H}_{q}$  may be written as
\cite{HJPW}:
\begin{equation}\label{h7}
e^{(N)}_{q} \simeq \frac{\hbar }{{q}^{2/3}} \sqrt{2 \frak{K}_2
\frak{K}_3} \, (N+1) \, .
\end{equation}
 The expression for $e^{(N)}_q$ is rather a
rough approximation for large values of $N$, since
$V(\hat{\beta})$ is highly nonharmonic far away from its minimum.
But for small values of $N$, this expression is valid at any value
of ${q}$.

\noindent{\it  Validity of approximation.} We notice in Eq.\;\eqref{h7} 
that the discrete spectrum part in Eq.\;\eqref{h6} becomes a small perturbation 
at large $q$, a range for which BOA possibly loses its validity,
whereas it gains all its value at small $q$. From the mass
criterion, our approach based on  BOA is legitimate as $q$
assumes its values near the singularity $q = 0$. Furthermore,
 our procedure of
quantization generates a supplementary repulsive potential that
prohibits the system to access the singularity neighborhood $q \in (0,q_m)$ with
some very small bound $q_m>0$, which depends on the initial state.

Furthermore, calculations made in
molecular physics beyond  BOA (the so-called vibronic
approximation) show that the mass criterion is in fact too strong:
a significant breakdown of BOA only occurs when different
eigenenergy curves $q \mapsto e^{(N)}_q$ of
$\hat{\mathcal H}_q $ are crossing. In our approach these
crossings do not occur, at least for the lowest levels of
$\hat{\mathcal H}_q $ in Eq.\,\eqref{h7}.

This reasoning based on molecular physics is robust, but qualitative
in our case, due to the coupling between the $q$ and $\beta_\pm$
degrees of freedom which is not of molecular type. In \cite{HJPW}
we weaken the adiabatic condition by allowing the quantized
oscillations to be excited by the semiclassical dynamics of the
isotropic background described below, for a fixed $N$. We find
that the excitation is indeed very limited, which justifies our
approach.

\noindent{\it   Semiclassical Hamiltonian.} We introduce a
semiclassical observable associated with the quantum Hamiltonian
(\ref{h6}) as its expectation value $\check{\mathcal{C}}_A (q,p):=
\langle  q, p | \hat{\mathcal{C}}_A |q,p\rangle$ in the ACS state
$| q, p \rangle$ peaked on the classical phase space point $(q,p)$
in the half-plane,
\begin{equation*}\label{h77}
\check{\mathcal{C}}_A (q,p) =\frac{3}{16} \left( p^2 +
\frac{\hbar^2 \frak{K}_4}{q^2} \right) +\frac{3}{4}
\frak{K}_5 q^{2/3}  - \frac{\hbar }{q^{2/3}} \frak{K}_6 (N+1)
 \, ,
\end{equation*}
where  the $\frak{K}_i$'s  are  positive numerical constants
\cite{HJPW} which are also simple rational functions of modified
Bessel functions $K_l(\nu)$. With our choice of $|\nu\rangle$,  at
large $\nu$, $\frak{K}_i(\nu) \sim 1$, $i\neq 4$ and
$\frak{K}_4(\nu) \sim \nu/2$. For the consistency of our
semiclassical description we have rescaled the fiducial vector so
that $\langle q, p | \hat{q}| q, p \rangle = q$ and $\langle q, p
| \hat{p}| q, p \rangle = p$.

The Hamiltonian constraint imposed at the semiclassical level,
$\check{\mathcal{C}}_A(q,p)=0$, leads to the semiclassical Friedmann-like equation:
\begin{equation}\label{c1}
H^2 + \frac{4 \pi^2 G^2 \hbar^2}{ c^{4}}
\frac{\frak{K}_4}{a^6}+\frac{\frak{K}_5}{4}
\Big(\frac{c}{a}\Big)^2 = \frac{8 \pi G\hbar}{3 c} (N+1) \frac{
\frak{K}_6}{ a^4}  \, .
\end{equation}
where we have restored physical constants and the standard
cosmological variables.

\begin{figure}
\includegraphics[width=6cm]{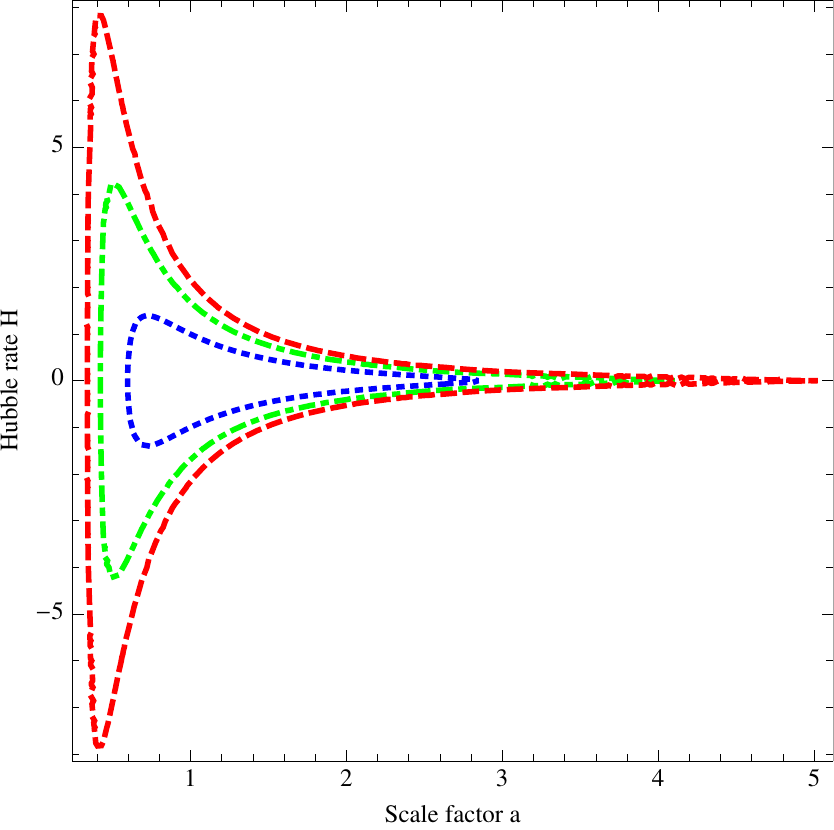}
\caption{Three periodic semiclassical trajectories in the half-plane
$(a,H)$ from Eq.\,\eqref{c1}. They are smooth plane curves. We use standard units $8 \pi
G=c=\hbar=1$ and choose $\nu=1$. Blue dotted curve for $N=0$,
green dotdashed for $N=1$ and red dashed for
$N=2$.}\label{figure3}
\end{figure}

The above semi-classical constraint  admits smooth trajectories
for all values of $N$ only if $\nu \in (0, 7.19)$. For $\nu
>7.19$, Eq. \eqref{c1} has no solution for the smallest values of
$N$. The solution of  \eqref{c1}  for $a$ is a periodic function,
$a \in [a_-,a_+]$ with $a_->0$ and $a_+  <  \infty$, and resolves
the cosmological singularity of the Mixmaster universe. In Fig. 2
we plot a few trajectories in the half-plane $(a,H)$. The
classical closed FRW model is recovered at $\hbar = 0$ and large
values of $\nu$.

\noindent{\it  Discussion.} Our semi-classical analysis of the
Mixmaster universe leads  to the modified Friedmann equation
\eqref{c1}. The left-hand side describes the isotropic part of
geometry. The Hubble parameter squared is accompanied by the
repulsive potential of purely quantum origin, which grows as
$a^{-6}$ during the contraction. At small volumes, it efficiently
counteracts the attraction of common forms of matter, forcing the
collapsing universe to rebound. The third term is the usual
isotropic spatial curvature.

The right-hand side of Eq.\;\eqref{c1} describes the quantized
energy of the anisotropy oscillations. The energy is discrete and
increases linearly with integer $N$, as expected in our harmonic
approximation. Within the adiabatic approximation, that quantum
number is conserved during the evolution. The
energy of the quantum oscillator evolves due to the $a$-dependent coefficients in front of its kinetic and potential terms given in Eq.\;\eqref{h5}. The ratio between the coefficients determines the oscillator's frequency,
which is proportional to $a^2$. The energy of the oscillations at the quantum level is multiplied by the frequency and consequently it
scales as $ a^{-4}$. (This   becomes a poor approximation for high values
of $N$, due to the breakdown of the harmonic approximation).  It is quite a contrast
to the classical wave, whose total energy is approximately
unaffected by its time-dependent frequency, and therefore scales
as $ a^{-6}$. Thus, the growth of the attractive force induced by
the anisotropy is significantly reduced in the semiclassical
dynamics. This essential dissimilarity between the classical and
semiclassical dynamics is due to the fact that on the quantum
level the contraction of space is driven by a quantum average.

Let us note that the energy of the wave does not vanish even in
the ground state $N=0$ due to the zero-point quantum fluctuations
corresponding to the classical state $\beta_\pm=p_\pm=0$. In
\cite{HJPW} we go beyond the adiabatic approximation to check if
there is a significant excitation of the wave's energy level
during the semiclassical evolution of the background geometry. The
method is essentially the same as the one  used to discuss the
generation of primordial power spectra in inflationary cosmology.
We show that the  wave in fact remains in its lowest energy states
during the quantum phase. It confirms that the quantum  FRW
universe, unlike its classical version, is dynamically stable with
respect to the small isotropy perturbation. Therefore, it seems
that the quantum closed Friedmann model may be successfully used to
describe also the earliest Universe, provided that the corresponding
Hamiltonian is supplemented with the effect of the zero-point
energy generated by the quantized anisotropy degrees of freedom of
the Mixmaster universe.\\

\acknowledgements
We thank Nathalie Deruelle and Martin Bucher for their remarks. P.M. was supported by MNiSW Fellowship ``Mobilno\'s\'c Plus".

\end{document}